\documentclass[12pt]{article}

\usepackage[T1]{fontenc}
\usepackage[utf8]{inputenc}
\usepackage{eufrak,amsmath,amssymb,accents,mathrsfs}
\usepackage[retainorgcmds]{IEEEtrantools}
\usepackage{multirow}
\usepackage{multicol}
\usepackage{tikz}
\usepackage{graphicx,color}
\usepackage{booktabs}
\usepackage{placeins}
\usepackage{mathtools}
\usepackage{calc}
\usepackage[nosort,nomove,compress]{cite}
\usepackage[scr=boondox,frak=euler,scrscaled=1.10,frakscaled=1.20]{mathalpha}
\usepackage{bm}
\usepackage[scaled=0.9]{helvet}
\usepackage[letterpaper,
			hmargin={1.9cm,1.9cm},
			vmargin={1.9cm,1.9cm},
			footskip=8mm]{geometry}
\usepackage{setspace}
\usepackage[colorlinks,
			breaklinks,
			hyperfootnotes,
			linktoc=page,
            linkcolor=blue,
            citecolor=blue,
            urlcolor=blue,
            unicode,
            pdfstartview=FitH,
            bookmarks,
            bookmarksnumbered]{hyperref}

\renewcommand\a {{\alpha}}
\renewcommand\b {{\beta}}
\renewcommand\r {{\rho}}

\renewcommand\th {{\theta}}

\newcommand\ad {{\dot{\alpha}}}
\newcommand\bd {{\dot{\beta}}}
\newcommand\gd {{\dot{\gamma}}}

\newcommand\thd {{\bar{\theta}}}

\newcommand\N{{\mathcal{N}}}

\newcommand\s{{\mathbscr{s}}}
\renewcommand\S{{\mathfrak{s}}}
\newcommand\g{{\bm{\gamma}}}
\renewcommand\d{{\bm{\delta}}}
\newcommand\Q{{\bm{Q}}}
\renewcommand\c{{\bm{c}}}
\newcommand\rb{{\bm{\r}}}
\newcommand\bb{{\bm{\b}}}

\newcommand\D {{\rm D}}
\newcommand\Dd {{\bar{\rm D}}}
\newcommand\pa {{\partial}}

\def\bea{\begin{IEEEeqnarray*}}
\def\eea{\end{IEEEeqnarray*}}
\def\be{\begin{eqnarray}}
\def\ee{\end{eqnarray}}
\def\n{\IEEEyesnumber}
\def\sn{\IEEEyessubnumber}

\allowdisplaybreaks

\makeatletter
\renewcommand\section{\@startsection{section}{1}{\z@}
              {3ex plus-1ex minus-.2ex}{1pt plus1pt}
              {\large\sf\bfseries\boldmath}}
\renewcommand{\subsection}{\@startsection{subsection}{2}{\z@}
              {1.5ex plus-1ex minus-.2ex}{0.01pt plus1pt}{\sf\slshape}}
\renewcommand{\subsubsection}{\@startsection{subsubsection}{3}{\z@}
              {1.5ex plus-1ex minus-.2ex}{0.01pt plus0.2pt}{\sf\boldmath}}
\renewcommand{\paragraph}{\@startsection{paragraph}{4}{\z@}
              {.75ex \@plus.5ex \@minus.2ex}{-2mm}{\sf\bfseries\boldmath}}
\makeatother

\newcommand{\Title}[1]{ {\large \bf #1 \vspace{3ex}} }
\newcommand{\Author}[3]{ {\large #1\footnote{\href{mailto:#2}{#2}}$^{#3}$} }
\newcommand{\Inst}[2]{ \emph{\centering $^{#2}$#1} }
\newcommand{\Abstract}[1]{ { ABSTRACT}\\ [4mm]
  \parbox{142mm}{\parindent=2pc\indent\baselineskip=14pt plus1pt #1}}

\begin{document}
\thispagestyle{empty}
\vspace*{6mm}
\begin{center}
    \Title{Superspace BRST/BV operators of superfield gauge theories} \\   [9mm]
\Author{I.~L.~Buchbinder}{joseph@tspu.edu.ru}{a,b},~
\Author{S.~James Gates Jr.}{sylvester\_gates@brown.edu}{c,d},~
\Author{K.~Koutrolikos}{konstantinos\_koutrolikos@brown.edu}{c,d}\\ [8mm]
\Inst{Center of Theoretical Physics,Tomsk State Pedagogical University,\\
      5 Tomsk 634041, Russia}{a}\\[8pt]
\Inst{Laboratory for Theoretical Cosmology, International Center for Gravity and Cosmos,\\
    Tomsk State University of Control Systems and Radioelectronics (TUSUR),\\
634050, Tomsk, Russia}{b}\\[8pt]
\Inst{Brown Theoretical Physics Center,\\[1pt]
      Box S, 340 Brook Street, Barus Hall,
      Providence, RI 02912, USA}{c}\\[8pt]
\Inst{Department of Physics, Brown University,\\[1pt]
      Box 1843, 182 Hope Street, Barus \& Holley 545,
      Providence, RI 02912, USA}{d}\\ [10mm]
\Abstract{
We consider the superspace BRST and BV description of $4D,~\N=1$ Super Maxwell theory and its non-abelian
generalization Super Yang-Mills. By fermionizing the superspace gauge transformation of the gauge superfields
we define the nilpotent superspace BRST symmetry transformation ($\s$). After introducing an appropriate set
of anti-superfields and define the superspace antibracket, we use it to construct the BV-BRST nilpotent
differential operator ($\S$) in terms of superspace covariant derivatives. The anti-superfield independent terms
of $\S$ provide a superspace generalization of the Koszul-Tate resolution
($\d$). In the linearized limit, the set of superspace differential operators that appear in $\S$ satisfy a
nonlinear algebra which can be used to construct a BRST charge $\Q$ without requiring pure spinor variables.
$\Q$ acts on the Hilbert space of superfield states and its cohomology generates the expected superspace
equations of motion.
}
\end{center}
\vfill
%
\clearpage
%
\section{Introduction}\label{sec:intro}
The Becchi-Rouet-Stora-Tutin (BRST) formalism offers a modern approach of describing gauge theories.  The
power of this technique has been demonstrated repeatedly from the formulation of field theories that
correspond to given first quantized systems all the way to string field theory and the study of string
interactions.

Initially, BRST symmetry \cite{Becchi:1974xu, Becchi:1974md, Becchi:1975nq, Tyutin:1975qk} was introduced as a
method of quantizing gauge field theories. In general, the quantization procedure of gauge theories is not
straightforward and requires the involvement of ghost fields. The two basic characteristics of gauge theories
that lead to the introduction of ghosts are (a) gauge redundancy and (b) the tensorial nature of gauge fields.
Specifically, one must eliminate the non-dynamical degrees of freedom of gauge fields and also eliminate the
states with non-positive definite norm which correspond to the time components\footnote{We use the mostly plus
convention.} of gauge fields. The first was completely understood by the Fadeev-Popov procedure from the view
point of a path-integral integration measure effect. Both these issues are addressed in the BRST quantization
procedure which systematically generates not only the Fadeev-Popov ghosts ---which give rise to appropriate
gauge fixing conditions--- but also additional propagating ghosts (non-minimal sector) that cancel the
negative norm states.

This approach was later extended to the Batalin-Vilkovisky (BV)\cite{Batalin:1981jr, Batalin:1983ggl,
Batalin:1985qj} formalism which promotes the above BRST symmetry to a fundamental principal of the theory
which is automatically incorporated by the introduction of additional fields, the antifields. The antifields
have the interpretation of sources of BRST transformations and were initially introduced as a technique to
capture the renormalization of the composite operators emerged in the BRST transformations of interacting
gauge theories. However, antifields play a crucial role whether one studies quantum or classical aspects of
the theory. They allow the definition of a symplectic structure in the space of fields and antifields,
the antibracket. The antibracket can be understood as a covariant analog of the Hamiltonian Poisson bracket
and as a consequence various Hamiltonian concepts (e.g. canonical transformations) can be introduced and
applied\cite{Voronov:1982ph, Voronov:1982cp, Voronov:1982ur}. This makes the field-antifield formalism a very
powerful tool which can be used in cases where the usual Fadeev-Popov method may fail. An important class
of such theories are the ones with open symmetry algebras, meaning that the commutator of two symmetries
generates trivial symmetries. In general, supersymmetric theories fall in this class, unless one considers
theories with off-shell supersymmetry, like the ones that have a superspace description\footnote{See the
relevant discussion in \cite{Buchbinder:2021igw}.}.

Moreover, the antibracket formalism gained popularity among string theorists when it was applied to the
bosonic open string field theory\cite{Witten:1985cc,Bochicchio:1986bd, Thorn:1986qj, Thorn:1988hm} and later
to closed string field theory\cite{Zwiebach:1992ie}.  This proved to be very useful for the study of string
interactions and various string vacua containing non-perturbative objects like D-branes\cite{Taylor:2003gn}.
The BRST approach to string field theory originated in the work of Siegel\cite{Siegel:1988yz} where a
nilpotent, fermionic BRST operator is constructed which commutes with observables, acts on the state space of
the theory and its cohomology defines the physical spectrum. This state space BRST operator is a reflection of
the BRST symmetry of the target space fields as generated from the antibracket in the Lagrangian path integral
formulation.  Interestingly, similar state space BRST operators have been constructed to describe higher spin
gauge fields \cite{Bekaert:2003uc, Buchbinder:2004gp, Buchbinder:2005ua, Buchbinder:2006eq, Bekaert:2006us,
Buchbinder:2007vq, Polyakov:2009pk, Henneaux:2012wg, Metsaev:2012uy, Henneaux:2013gba, Buchbinder:2015kca,
Buchbinder:2018yoo, Buchbinder:2021qrg, Buchbinder:2021rmy, Buchbinder:2021rfm, Buchbinder:2021xbk}.  This is
another indication of the close relation between higher spins and string theory.

Finally, for maximally supersymmetric theories, it has been shown that a covariant quantization can be
achieved by using pure spinor variables\cite{Berkovits:2000fe} and the spectrum of these theories is captured
by the cohomology of a pure spinor BRST operator\cite{Cederwall:2001dx}.  The free action was
constructed\cite{Berkovits:2002zk} in terms of a pure spinor superfield\footnote{A superfield with an
additional dependence on a pure spinor variable $\lambda$} $\Psi(x,\theta,\lambda)$ and the pure spinor BRST
operator $Q=\lambda^{\a}\D_{\a}$.  However, these descriptions were not manifestly maximally supersymmetric
and only worked for linearized theories. Later, both issues were addressed by the introduction of additional
--non-minimal-- pure spinors that modified the BRST charge operator\cite{Berkovits:2005bt} and by adopting
the BV formalism and solving the master equation\cite{Cederwall:2010tn, Cederwall:2013vba}.

The purpose of this paper is to explore the BRST symmetry, BV and BFV (Batalin-Fradkin-Vilkovisky)
formulations of supersymmetric gauge theories\footnote{See \cite{Gates:1989hg, Gates:1990ze} for earlier
considerations.} from the superspace point of view without the use of pure spinor variables. Similarly to
string theory applications, we construct a superspace BRST operator acting on the (target) space of
superfields and a corresponding BRST operator that acts on a Hilbert space of superfield states.

Following the BRST procedure, for a given supersymmetric gauge theory with a superspace description we
introduce a nilpotent, BRST symmetry operator $\mathbscr{s}$ ($\s^2=0$) by fermionizing the superspace gauge
transformations of the theory via the replacement of all gauge parameter superfields with corresponding ghost
superfields of opposite statistics.  The superspace Lagrangian is deformed by the addition of an appropriate
$\mathbscr{s}$-exact term which includes additional (non-minimal sector) ghost superfields like the
Nakanishi-Lautrup ghost superfield and its BRST-doublet partner.

In the BV description, for every superfield and ghost superfield we introduce a conjugate variable, the
anti-superfield. In the superfield/anti-superfield space we define the superspace antibracket and use it to
construct a nilpotent, superspace differential operator $\mathfrak{s}$~($\S^2=0$). This is a superspace BRST
differential operator which action on (anti-) superfields can be decomposed into two pieces:
$\S=\bm{\gamma}+\bm{\delta}$. The $\bm{\gamma}$ part is the now renamed superspace BRST symmetry and its
action on superfields coincides with $\s$. The second part $\bm{\delta}$ is the superspace generalization of the
Koszul-Tate resolution\cite{Henneaux:1992ig}
which implements the superspace equations of motion. As expected both $\bm{\gamma}$ and $\bm{\delta}$ are
nilpotent and anticommute.

Finally, getting inspiration from string field theory, we define a Hilbert space of states where the arbitrary
state vector $|\bm{\Psi}\rangle$ can be expanded in some basis $|\bm{\phi_s}\rangle$ with superfield
coefficients $\bm{\chi_{s}}$ ($|\bm{\Psi}\rangle=\sum_s|\bm{\phi_s}\rangle\bm{\chi_s}$). In this Hilbert
space, we define a nilpotent BRST operator $\bm{Q}$~($\bm{Q}^2=0$), such that its action on the Hilbert space
vectors coincides with the action of $\S$ on the superfield coefficients. This is done in two steps: (1)
we identify the set of differential operators that appear in $\S$ and
(2) find their algebraic properties in order to construct a nilpotent Hilbert space BRST operator via
the Fradkin-Fradkina algorithm\cite{Fradkin:1977xi}.  The cohomology of $\Q$ must correctly
generate the superspace equations of motion and gauge transformations for the superfields of the theory and
thus produce the expected physical spectrum.

We would like to emphasize that although the existence of such BRST charge is proven for general gauge
theories\cite{Fradkin:1975cq, Batalin:1977pb, Batalin:1983pz, Henneaux:1985kr}, the explicit construction of
it can be a nontrivial task. For supersymmetric theories with a superspace description the only available
methodology for constructing the BRST charge is based on pure spinors. In the pure spinor approach, the
nilpotence of the BRST charge is based on the pure spinor constraint and not on the algebraic properties of
the characteristic set of differential operators that participate in the description of the theory. Our
methodology can offer an alternative method of constructing such nilpotent charges. This could be beneficial
for the construction of a manifestly supersymmetric BRST charge that describes higher spin supermultiplets.

The proposed procedure can be applied to any gauge theory in superspace. In this paper, we demonstrate it for
$4D,~\N=1$ Super Yang-Mills (SYM) and its linearized limit, the vector supermultiplet. We explicitly construct
the superspace BRST symmetry operator $\S$ and extract the superspace Koszul-Tate complex $\d$.  For the
linearized theory, the set of superspace differential operators that appear in $\S$ satisfy a nonlinear
algebra. Nevertheless, using known corrections\cite{Schoutens:1989tn, Buchbinder:2007au} to the
Fradkin-Fradkina algorithm we find and explicit expression for $\Q$ in terms of supersymmetric covariant
derivatives without the need to introduce pure spinors. The cohomology of this superspace BRST charge
correctly generates the gauge symmetry and superspace equations of motion for the vector supermultiplet.

Before discussing all the above, we first remind to the reader the BRST, BV and BFV descriptions in the context
of a simple example: Maxwell theory. This corresponds to the bosonic sector of the linearized SYM and it is
the simplest example where the non-supersymmetric shadows ($\mathscr{s},\mathtt{s},\gamma,\delta,Q$) of the
quantities discussed above can be introduced and easily constructed.  Moreover,  it is explicitly shown that
there is a Hilbert space of field states and a nilpotent charge $Q$ such that the action of the BRST operator
$\mathtt{s}$, generated by the antibracket, on fields and antifields components of the field states, is equal
to the action of $Q$ on the basis vectors of the Hilbert state $\{|e_i\rangle\}$:~
$\mathtt{s}|\psi\rangle=Q|\psi\rangle$.  The BRST charge $Q$ constructed this way is identical to the
Batalin-Fradkin-Vilkovisky (BFV) nilpotent charge constructed from the algebra of the physical constraints that
define the on-shell propagating degrees of freedom of the theory.
\section{Review of BRST, BV and BFV descriptions of Maxwell theory}
In the Lagrangian path integral approach, the BRST method is based on the fact that if the theory under
consideration is a gauge theory, then there is a nilpotent, odd BRST symmetry ($\mathscr{s}$) and one can
deform the Lagrangian of the theory by the addition of an $\mathscr{s}$-exact term without affecting the path
integral
\begin{equation}\label{Omega}
    \mathcal{L}\to\mathcal{L}+\mathscr{s}\Omega~.
\end{equation}
The deformed action remains invariant
\begin{equation}
    \mathscr{s}\mathcal{L}+\mathscr{s}^2\Omega=0
\end{equation}
because both terms independently vanish. The second term vanishes because $\mathscr{s}$ must be nilpotent
($\mathscr{s}^2=0$) and the first term vanishes because we choose the BRST transformation of the fields that
appear in the starting Lagrangian $\mathcal{L}$ to have exactly the same form as their corresponding gauge
transformations. Therefore, the Bianchi identities that guarantee the gauge invariance of $\mathcal{L}$ will
also guarantee $\mathscr{s}$-invariance. The tension between $\mathscr{s}$ being an odd symmetry and also
having the same structure as the gauge transformations is resolved by the replacement of every gauge parameter
by a ghost field of opposite statistics. This way one can immediately define the BRST symmetry
transformation of any gauge field.  This is usually referred to as fermionization of the gauge transformation.
For the case of Maxwell's theory, we get:
\begin{equation}\label{sA}
    \delta A_{m}=\pa_m\lambda~\to~\mathscr{s}A_m=\pa_m c
\end{equation}
where $c$ is a ghost field. Moreover, because of the nilpotence of this BRST transformation, one
determines the BRST transformation of the ghost field:
\begin{equation}\label{sc}
    \mathscr{s}^2A_m=\pa_m(\mathscr{s}c)=0~\Rightarrow~\mathscr{s}c=0~.
\end{equation}
Because the deformation \eqref{Omega} of the Lagrangian is $\mathscr{s}$-exact, the theory is independent of
any particular choice of $\Omega$. In the context of quantizing the theory one selects this deformation such
that it generates an appropriate gauge fixing condition by integrating out a ghost field, as expected from
the Fadeev-Popov procedure. For Maxwell's theory usually one chooses the Landau-Fermi gauge or more generally
the Feynman gauge:
\begin{equation}
    \mathcal{L}\to\mathcal{L}+\r(\pa^mA_m+\tfrac{\xi}{2}\r)
\end{equation}
where $\r$ is a bosonic ghost field and $\xi$ is the usual Feynman gauge parameter. However, this deformation
fails to satisfy the consistency condition of being $\mathscr{s}$-closed:
\begin{equation}
    \mathscr{s}[~\r(\pa^mA_m+\tfrac{\xi}{2}\r)~]=\mathscr{s}(\r)~(\pa^mA_m+\xi\r)+\r\Box c\neq0~.
\end{equation}
Even if we choose $\mathscr{s}\r=0$, there is no way to cancel the second term, hence in order for the
deformation to be consistent we must add additional terms. It is straightforward to check that the
complete answer is:
\begin{equation}
    \mathscr{s}\Omega=\r(\pa^mA_m+\tfrac{\xi}{2}\r)-\b\Box c
\end{equation}
where $\b$ is an additional fermionic ghost field. By assigning to $\r$ and $\b$ the
following $\mathscr{s}$ transformations
\begin{equation}\label{srsb}
    \mathscr{s}\r=0~,~\mathscr{s}\b=\r
\end{equation}
one can check that the right hand side not only is $\mathscr{s}$-closed but also $\mathscr{s}$-exact like the
left hand side, with $\Omega$ taking the form
\begin{equation}
    \Omega=\b(\pa^mA_m+\tfrac{\xi}{2}\r)~.
\end{equation}
The fermionic ghost fields $\b$ and $c$ correspond to the Fadeev-Popov ghosts and the bosonic ghost field
$\r$ is the Nakanishi-Lautrup field. Moreover, the ($\b,\r$) pair defines the \emph{non-minimal sector} of the
theory, which must have two properties. First, it must allow the construction of an action for the fermionic
ghost c (accomplished by $\b$). Second, $\b$ can not be BRST isolated ($\mathscr{s}$-closed) because it
may introduce new gauge invariant quantities and affect the physical sector of the theory
(accomplished by $\r$).
Transformations \eqref{srsb} have a special structure which identifies the non-minimal sector
ghosts $\b$ and $\r$ as a \emph{BRST-doublet}. BRST-doublets play a very crucial role in BRST constructions
because they do not contribute non-trivial pieces in the cohomology of the theory. In the case or string
theory or the (super)particle similar BRST-doublets are introduced.

In the BV formalism, one goes a step further and for every field / ghost introduce a ``\emph{conjugate}''
variable, the antifield. For the moment, antifields are considered as classical sources that generate
$\mathscr{s}$ transformations of the corresponding fields or ghosts and thus appear in the action in the
following manner: $\text{Antifield}~\mathscr{s}(\text{Field})$. Therefore the antifield of a specific field /
ghost carries appropriate quantum numbers / indices in order for these terms to be well defined additions to
the deformed Lagrangian. For Maxwell theory we get
\begin{equation}
    \mathcal{L}_{BV}=\tfrac{1}{2}A^m(\Box A_m-\pa_m\pa^nA_n)+\r(\pa^mA_m+\tfrac{\xi}{2}\r)-\b\Box c~+
    A^{*m}\pa_m c+\b^{*}\r
\end{equation}
where $A^{*m}, c^{*}, \r^{*}, \b^{*}$ are the antifields of $A_m, c, \r, \b$ respectively. In this case,
$c^{*}$ and $\r^{*}$ drop out of the Lagrangian because their corresponding ghost fields are
$\mathscr{s}$-closed. Furthermore, in order for these additional terms to be BRST invariant:
\begin{equation}
    \mathscr{s}[\text{Antifield}~\mathscr{s}(\text{Field})]=
    \mathscr{s}(\text{Antifield})~\mathscr{s}(\text{Field})=0~\Rightarrow~\mathscr{s}\text{Antifield}=0
\end{equation}
we choose all antifields to be $\mathscr{s}$-closed.

For a general gauge theory, the variation of the BV action ($S$) under the $\mathscr{s}$-BRST transformations
takes the form:
\begin{equation}
    \delta S=\int~\Big\{\mathscr{s}\Phi^{i}\frac{\delta S}{\delta\Phi^{i}}+\mathscr{s}\Phi^{*}_{i}
    \frac{\delta S}{\delta\Phi^{*}_{i}}\Big\}
\end{equation}
where $\Phi^{i}$ is symbolically the set of gauge fields and ghosts that participate in the BRST description
of the gauge theory as discussed above and $\Phi^{*}_{i}$ are their corresponding antifields. However, because
the antifields are $\mathscr{s}$-closed and also appear in the action such that $\frac{\delta
S}{\Phi^{*}_{i}}=\mathscr{s}\Phi^{i}$ one arrives to the expression:
\begin{equation}
    \delta S=\int~\frac{\delta S}{\delta\Phi^{*}_{i}}\frac{\delta S}{\delta\Phi^{i}}~.
\end{equation}
This motivates the definition of a binary bracket,  called the antibracket
\begin{equation}
    \Big(F,G\Big)=\int~\Big\{\frac{\delta F}{\delta\Phi^{*}_{i}}\frac{\delta G}{\delta\Phi^{i}}+
    \frac{\delta G}{\delta\Phi^{*}_{i}}\frac{\delta F}{\delta\Phi^{i}}\Big\}~.
\end{equation}
Using the antibracket the $\mathscr{s}$-invariance of the action S takes the simple form
\begin{equation}
    \Big(S,S\Big)=0
\end{equation}
which is known as the \emph{classical master equation}. It is easy to check that the fields and antifields are
conjugate variables with respect to the antibracket but most importantly because the action obeys the
classical master equation one can define a nilpotent differential operator $\mathtt{s}~(\mathtt{s}^2=0)$
that acts in the space of fields and antifields
\begin{equation}\label{s}
    \mathtt{s}F\equiv \Big(S,F\Big)
\end{equation}
for any field or antifield $F$.
The $\mathtt{s}$ transformations of the fields and antifields of Maxwell theory are:
\begin{IEEEeqnarray*}{ll}\n\label{sMt}
    \mathtt{s}A_m=\pa_m c~,~&\mathtt{s}A^{*m}=\Box A^m-\pa^m\pa_nA^n-\pa^m\r\sn\\
    \mathtt{s}\r=0~,~&\mathtt{s}\r^{*}=\pa^m A_m+\r+\b^{*}\sn\\
    \mathtt{s}\b=\r~,~&\mathtt{s}\b^{*}=-\Box c\sn\\
    \mathtt{s}c=0~,~&\mathtt{s}c^{*}=\Box \b+\pa_mA^{*m}\sn
\end{IEEEeqnarray*}
These transformations are nilpotent as expected and they have very interesting structure. The
$\mathtt{s}$-transformation of any field matches exactly their $\mathscr{s}$ transformation, whereas the
$\mathtt{s}$-transformation of the antifields knows about the equations of motion of the corresponding fields.

In general, the differential operator $\mathtt{s}$ can be decomposed into two operators:
$\mathtt{s}=\gamma+\delta$. Operator $\gamma$ is the newly renamed BRST symmetry transformation and is
defined as follows:
\begin{equation}\label{g}
    \gamma{\Phi^{i}}\equiv\frac{\delta S}{\delta \Phi^{*}_{i}}=\mathscr{s}\Phi^{i}~,~
    \gamma{\Phi^{*}_{i}}\equiv\frac{\delta }{\delta \Phi^{i}}\Big[\Phi^{*}_{j}~\mathscr{s}(\Phi^{j})\Big]~.
\end{equation}
For fields, $\gamma$ coincides with the $\mathscr{s}$ transformation but for antifields it has additional
terms originating from the variation of the antifield terms in the action.
The $\delta$ part is called the Koszul-Tate resolution\footnote{See \cite{Henneaux:1992ig, Henneaux:1989jq}}
and it implements the equations of motion. It is defined as follows:
\begin{equation}\label{d}
    \delta\Phi^{i}\equiv 0~,~\delta\Phi^{*}_{i}\equiv\frac{\delta}{\delta\Phi^{i}}
    \Big[S-\Phi^{*}_{j}~\mathscr{s}(\Phi^{j})\Big]~.
\end{equation}
Adding equations \eqref{g} and \eqref{d} automatically gives $\mathtt{s}$ as defined in \eqref{s}.
Additionally, $\gamma$ and $\delta$  are nilpotent and anticommute: $\gamma^2=0$,
$\delta^2=0$, $\gamma\delta+\delta\gamma=0$.

For Maxwell's theory the action of $\gamma$ and $\delta$ operators on the fields and antifields is:
\begin{IEEEeqnarray*}{llll}\n
\gamma A_m=\pa_m c~,~&\gamma A^{*m}=0~,~&\delta A_m=0~,~&\delta A^{*m}=\Box A^{m}-\pa^m\pa_nA^n-\pa^m\r\sn\\
    \gamma \r=0~,~&\gamma \r^{*}=\b^{*}~,~&\delta \r=0~,~&\delta \r^{*}=\pa^m A_{m}+\r\sn\\
    \gamma \b=\r~,~&\gamma \b^{*}=0~,~&\delta \b=0~,~&\delta \b^{*}=-\Box c\sn\\
    \gamma c=0~,~&\gamma c^{*}=\pa_mA^{*m}~,~&\delta c=0~,~&\delta c^{*}=\Box \b\sn
\end{IEEEeqnarray*}

Having constructed the BRST operator $\mathtt{s}$ on the space of fields and antifields, we can define a
corresponding BRST charge operator $Q$ which acts on an appropriately defined Hilbert space of states.  For
every pair of field $\Phi^{i}$ and antifield $\Phi^{*}_{i}$ we introduce a corresponding pair of state vectors
$|e^{(\Phi^{i})}\rangle$ and $|e^{(\Phi^{*}_{i})}\rangle$ and we consider the vector space which is spanned by
the states $\big\{|e^{(\Phi^{i})}\rangle,~|e^{(\Phi^{*}_{i})}\rangle\big\}$. The most general state is written as
\begin{equation}
    |\psi\rangle=\sum_i\chi_i|e^{(\Phi^{i})}\rangle+\sum_i\upsilon^{i}|e^{(\Phi^{*}_{i})}\rangle
\end{equation}
where $\chi_i$ and $\upsilon^{i}$ are field components of $|\psi\rangle$. Furthermore, for each of the basis
vectors we assign a ghost number value which is defined to be opposite to the ghost number of the
corresponding field or antifield. Specifically, if the field theoretic ghost number value of field $\Phi^{i}$
is $Gh(\Phi^{i})=g^i$ then we define the vector space ghost number value of the corresponding vectors to be
\begin{equation}
    gh(|e^{(\Phi^{i})}\rangle)\equiv -g^i~,~gh(|e^{(\Phi^{*}_{i})}\rangle)\equiv 1+g^i~.
\end{equation}
For Maxwell theory we have the following set of states:
\begin{IEEEeqnarray*}{l}
    gh\#=0 : \Big\{|e^{(A)}\rangle,~|e^{(\r)}\rangle,~|e^{(\b^{*})}\rangle\Big\}\\
    gh\#=1 : \Big\{|e^{(A^{*})}\rangle,~|e^{(\r^{*})}\rangle,~|e^{(\b)}\rangle\Big\}\\
    gh\#=-1 : \Big\{|e^{(c)}\rangle\Big\}\\
    gh\#=2 : \Big\{|e^{(c^{*})}\rangle\Big\}
\end{IEEEeqnarray*}
This structure of states can be captured by a Hilbert space with a vacuum state  $|\omega\rangle$ and three
fermionic creation operators $\eta, \zeta, \pi_{\xi}$ with ghost values $gh(\eta)=gh(\zeta)=-gh(\pi_{\xi})=1$
and $gh(|\omega\rangle)=0$
\begin{IEEEeqnarray*}{l}
    gh\#=0 : \Big\{|\omega\rangle,~\eta\pi_{\xi}|\omega\rangle,~\zeta\pi_{\xi}|\omega\rangle\Big\}\\
    gh\#=1 : \Big\{\eta|\omega\rangle,~\zeta|\omega\rangle,~\eta\zeta\pi_{\xi}|\omega\rangle\Big\}\\
    gh\#=-1 : \Big\{\pi_{\xi}|\omega\rangle\Big\}\\
    gh\#=2 : \Big\{\eta\zeta|\omega\rangle\Big\}
\end{IEEEeqnarray*}
In this basis, the most general state is a linear combination of the above states
\begin{equation}\label{psiexpand}
    |\psi\rangle=|\psi^{(-1)}\rangle+|\psi^{(0)}\rangle+|\psi^{(1)}\rangle+|\psi^{(2)}\rangle
\end{equation}
where
\begin{IEEEeqnarray*}{l}\n
    |\psi^{(-1)}\rangle=\pi_{\xi}|u\rangle~,~|u\rangle=u|\omega\rangle\sn\\
    |\psi^{(0)}\rangle=|w_1\rangle+\eta\pi_{\xi}|w_2\rangle+\zeta\pi_{\xi}|w_3\rangle~,~
    |w_i\rangle=w_i|\omega\rangle\sn\\
    |\psi^{(1)}\rangle=\eta|z_1\rangle+\zeta|z_2\rangle+\eta\zeta\pi_{\xi}|z_3\rangle~,~
    |z_i\rangle=z_i|\omega\rangle\sn\\
    |\psi^{(2)}\rangle=\eta\zeta|v\rangle~,~|v\rangle=v|\omega\rangle\sn
\end{IEEEeqnarray*}
and the coefficients $u, w_i, z_i, v$ are elements of the field-antifield space. Therefore the BRST operator
$\mathtt{s}$ can act on the state $|\psi\rangle$ by acting on these coefficients. On the other hand, we can
define a corresponding BRST operator $Q$ that acts on the states of this Hilbert space  which is equivalent to
the action of $\mathtt{s}$ \eqref{sMt}
\begin{equation}\label{Qs}
    Q|\psi\rangle=\mathtt{s}|\psi\rangle~.
\end{equation}
$Q$ must be odd, have ghost value one and must be constructed out of the oscillators of the Hilbert space
$\eta, \zeta, \pi_{\xi}$ and their conjugate $\pi_{\eta}, \pi_{\zeta}, \xi$:
$\{\eta,\pi_{\eta}\}=1,\{\zeta,\pi_{\zeta}\}=1,\{\xi,\pi_{\xi}\}=1$. The most general ansatz for $Q$ is
\begin{IEEEeqnarray*}{ll}\n
    Q=&~\eta\Lambda^{(0)}+\zeta\Lambda^{(1)}+\xi\Lambda^{(-1)}\\
      &+\eta\zeta\pi_{\eta}B_1+\eta\zeta\pi_{\zeta}B_2+\eta\zeta\pi_{\xi}B_3\\
      &+\eta\xi\pi_{\eta}\Gamma_1+\eta\xi\pi_{\zeta}\Gamma_2+\eta\xi\pi_{\xi}\Gamma_3\\
      &+\zeta\xi\pi_{\eta}\Delta_1+\zeta\xi\pi_{\zeta}\Delta_2+\zeta\xi\pi_{\xi}\Delta_3\\
      &+\eta\zeta\xi\pi_{\eta}\pi_{\zeta}K_1+\eta\zeta\xi\pi_{\eta}\pi_{\xi}K_2+
      \eta\zeta\xi\pi_{\zeta}\pi_{\xi}K_3\\
\end{IEEEeqnarray*}
Equation \eqref{Qs} can be expanded according to \eqref{psiexpand}
\begin{equation}
    Q|\psi^{(-1)}\rangle+Q|\psi^{(0)}\rangle+Q|\psi^{(1)}\rangle+Q|\psi^{(2)}\rangle=
    \mathtt{s}|\psi^{(-1)}\rangle+\mathtt{s}|\psi^{(0)}\rangle+\mathtt{s}|\psi^{(1)}\rangle
    +\mathtt{s}|\psi^{(2)}\rangle
\end{equation}
and by matching the Hilbert space ghost number (and the field-antifield space ghost number)
of the two sides we get the following set of equations
\begin{IEEEeqnarray*}{l}\n
    Q|\psi^{(-1)}\rangle=\mathtt{s}|\psi^{(0)}\rangle~,\sn\\
    Q|\psi^{(0)}\rangle=\mathtt{s}|\psi^{(1)}\rangle~,\sn\\
    Q|\psi^{(1)}\rangle=\mathtt{s}|\psi^{(2)}\rangle~,\sn\\
    Q|\psi^{(2)}\rangle=0=\mathtt{s}|\psi^{(-1)}\rangle~.\sn
\end{IEEEeqnarray*}
These equations can be solved to fix all the coefficients of $|\psi\rangle$ and determine all the field
operators in $Q$. We find:
\begin{IEEEeqnarray*}{ll}
    u=c~,~|u\rangle=|c\rangle=c|\omega\rangle~,&
    v=c^{*}~,~|v\rangle=|c^{*}\rangle=c^{*}|\omega\rangle~,\\
    w_1=A^{m}a^{\dagger}_{m}~,~|w_1\rangle=|A\rangle=A^{m}a^{\dagger}_{m}|\omega\rangle~,~&
    z_1=-A^{*m}a^{\dagger}_{m}-\pa^{m}a^{\dagger}_{m}~\r^{*}~,~
    |z_1\rangle=-|A^{*}\rangle-\Lambda^{(-1)}|\r^{*}\rangle~,~~~\\
    w_2=-\b^{*}~,~|w_2\rangle=-|\b^{*}\rangle=-\b^{*}|\omega\rangle~,~&
    z_2=-\r^{*}+(1+\pa^m a^{\dagger}_m)\b~,~|z_2\rangle=-|\r^{*}\rangle+(1+\Lambda^{(-1)})|\b\rangle~,~\\
    w_3=\r~,~|w_3\rangle=|\r\rangle=\r|\omega\rangle~,&
    z_3=-\Box \b~,~|z_3\rangle=-\Box|\b\rangle
    \end{IEEEeqnarray*}
where $a^{\dagger}_{m}$ and $a_{m}$ is a pair of bosonic creation and annihilation operators
($[a_m,a^{\dagger}_n]=\eta_{mn}$) and
\begin{equation}\label{Q}
    Q=\eta~\Box+\zeta~\pa^ma_m+\xi~\pa^ma^\dagger_m-\zeta\xi\pi_{\eta}~.
\end{equation}
This $Q$ operator is nilpotent by construction and it is identical to the BFV-BRST charge constructed
from the algebra of
operators $\Lambda^{(0)}=\Box,~\Lambda^{(1)}=\pa^ma_m,~\Lambda^{(-1)}=\pa^ma^{\dagger}_m$:
\begin{equation}
    [\Lambda^{(0)},\Lambda^{(1)}]=0~,~[\Lambda^{(0)},\Lambda^{(-1)}]=0~,~
    [\Lambda^{(1)},\Lambda^{(-1)}]=\Lambda^{(0)}~.
\end{equation}
The general physical state (zero ghost number) cohomology of Q is given by:
\begin{equation}
    |\phi\rangle=|A\rangle+\eta\pi_{\xi}|B\rangle+\zeta\pi_{\xi}|C\rangle
\end{equation}
with a transformation:
\begin{equation}\label{dspin1}
    \delta|\phi\rangle=Q|L\rangle~,~|L\rangle=\pi_{\xi}|\lambda\rangle~\Rightarrow~
    \begin{cases}
        \delta|A\rangle=\Lambda^{(-1)}|\lambda\rangle~,\\
        \delta|B\rangle=\Lambda^{(0)}|\lambda\rangle~,\\
        \delta|C\rangle=\Lambda^{(1)}|\lambda\rangle
    \end{cases}
\end{equation}
and equations of motion:
\begin{equation}\label{spin1}
Q|\phi\rangle=0~\Rightarrow~
    \begin{cases}
\Lambda^{(0)}|A\rangle-\Lambda^{(-1)}\Lambda^{(1)}|A\rangle+\Lambda^{(-1)}\Lambda^{(-1)}|C\rangle=0~,\\
\Lambda^{(1)}|A\rangle-\Lambda^{(-1)}|D\rangle-|B\rangle=0~,\\
\Lambda^{(0)}|D\rangle-\Lambda^{(1)}\Lambda^{(1)}|A\rangle+\Lambda^{(1)}\Lambda^{(-1)}|D\rangle=0~.\\
    \end{cases}
\end{equation}
For the case $|\phi\rangle=|\psi^{(0)}\rangle$ and $|L\rangle=|\psi^{(-1)}\rangle$ one recovers the
Maxwell theory equations and transformations.
\section{Superspace BRST description of Super Maxwell theory}\label{vs}
In this section, we apply the above concepts in superspace aiming towards a BRST-BV/BFV description
of the vector supermultiplet which is the supersymmetric extension of Maxwell theory.
The superspace description of the theory is given in terms of a real scalar superfield $V(x,\th,\thd)$
with the following superspace action principle
\begin{equation}\label{V}
    S=\tfrac{1}{2}\int d^4x d^2\th d^2\thd ~V\D^{\g}\Dd^2\D_{\g}V
\end{equation}
which is invariant under the gauge transformation $\delta V=\Dd^2L+\D^2\bar{L}$. The operators $\D_{\a}$ and
$\Dd_{\ad}$ are the supersymmetric covariant derivatives and the off-shell spectrum of the theory is a Maxwell
spin 1 gauge field, a decoupled auxiliary real scalar field and a spin 1/2 fermion.

Similarly to the discussion in the previous section, we promote the gauge symmetry of the theory to a
nilpotent, superspace BRST symmetry $\s$. The action of $\s$ on the gauge
superfield $V$ is found by fermionizing the original gauge transformation and replace the gauge parameter
superfields with ghost superfields:
\begin{equation}
    \s V=\Dd^2\c+\D^2\bar{\c}~.
\end{equation}
Using the nilpotence of $\s$ we also find that
\begin{equation}
    0=\s^2V=\Dd^2(\s\c)+\D^2(\s\bar{\c})~\Rightarrow~\s\c=0~,~\s\bar{\c}=0
\end{equation}
An important comment is that, unlike Maxwell theory, the differential operators that appears in the
gauge transformation of the gauge superfield have a non-zero kernel and based on the algebra of the
supersymmetric covariant derivatives we can introduce  ghosts for ghosts, ghosts for ghosts for ghosts
et cetera ad infinitum\footnote{
$\s\c=0+\Dd^{\ad}\bar{\bm{d}}_{\ad},~
\s\bar{\bm{d}}_{\ad}=0+\Dd^{\bd}\bar{\bm{d}}_{\ad\bd},~\s\bar{\bm{d}}_{\ad\bd}=0+
\Dd^{\gd}\bar{\bm{d}}_{\ad\bd\gd},\dots$ . Similarly for $\bar{\c}$.}. For the
purpose of this paper these contributions are not required and we will not consider them.

Next, we deform the Lagrangian by adding an appropriate $\s$-exact term, $\s\bm{\Omega}$. Motivated from the
quantization procedure for this theory, this deformation will include a gauge fixing condition accompanied by
the corresponding Lagrange multiplier ghost superfield. The usual superspace gauge fixing conditions for gauge
superfield $V$ are $\D^2V=0$ and $\Dd^2V=0$, which correspond to the superspace extension of the Landau-Fermi
gauge in Maxwell theory. Therefore, we consider the deformation:
\begin{equation}
    \s\bm{\Omega}=\rb(\D^2V+\tfrac{\xi}{2}\bar{\rb})+\bar{\rb}(\Dd^2V+\tfrac{\xi}{2}\rb)+\dots
\end{equation}
where $\rb$ and $\bar{\rb}$ are Nakanishi-Lautrup superfield ghosts, $\xi$ is the Feynman gauge parameter and
the dots represent additional terms that we have in order to make the right hand side of the equation
$\s$-exact.
Using the nilpotence of $\s$ we find the consistence condition
\begin{equation}
    0=(\s\rb)(\D^2V+\xi\bar{\rb})+(\s\bar{\rb})(\Dd^2V+\xi\rb)+\rb\D^2\Dd^2\c+\bar{\rb}\Dd^2\D^2\bar{\c}
    +\s(\dots)
\end{equation}
which determines the missing terms to be
\begin{equation}\label{superomega}
    \s\bm{\Omega}=\rb(\D^2V+\tfrac{\xi}{2}\bar{\rb})+\bar{\rb}(\Dd^2V+\tfrac{\xi}{2}\rb)
    -\bb\D^2\Dd^2\c -\bar{\bb}\Dd^2\D^2\bar{\c}~.
\end{equation}
The remaining $\s$ transformations are
\begin{equation}
    \s\bm{\b}=\rb~,~\s\rb=0~,~\s\bar{\bb}=\bar{\rb}~,~\s\bar{\rb}=0
\end{equation}
and $\bb$ is a fermionic ghost superfield which forms a BRST-doublet with $\rb$. Similarly for
$\bar{\bb}$ and $\bar{\rb}$. This is expected because there are two gauge parameter superfields $(L,\bar{L})$
and therefore by fermionizing them we get two fermionic ghost superfields $(\c,\bar{\c})$ and for each one of
them we must introduce a BRST-doublet, as discussed in the previous section.  The non-minimal sector of this
theory is the two BRST-doublets ($\bb,\rb$) and ($\bar{\bb},\bar{\rb}$).
Also, using the above $\s$ transformations one can check that the right hand side of \eqref{superomega} is
$\s$-exact and solve for $\bm{\Omega}$
\begin{equation}
    \bm{\Omega}=\bb(\D^2V+\tfrac{\xi}{2}\bar{\rb})+\bar{\bb}(\Dd^2V+\tfrac{\xi}{2}\rb)~.
\end{equation}

This deformation is further extended by the BV anti-superfields. For each (ghost) superfield we introduce
an anti-superfield with opposite statistics and appropriate superspace ghost number
and add to the action a $[\text{Anti-superfield}~\s(\text{Superfield})]$ term. The superspace BV
Lagrangian takes the form:
\begin{IEEEeqnarray*}{ll}
    \mathcal{L}_{BV}=&~\tfrac{1}{2}V\D^\gamma\Dd^2\D_\gamma V+\rb(\D^2V+\tfrac{\xi}{2}\bar{\rb})+
    \bar{\rb}(\Dd^2V+\tfrac{\xi}{2}\rb)-\bb\D^2\Dd^2\c -\bar{\bb}\Dd^2\D^2\bar{\c}\\
                     &+\bm{V}^{*}(\Dd^2\c+\D^2\bar{\c})+\bb^{*}\rb+\bar{\bb}^{*}\bar{\rb}
\end{IEEEeqnarray*}
where $\bm{V}^{*},~\bb^{*},~\bar{\bb}^{*},~\rb^{*},~\bar{\rb}^{*},~\c^{*},~\bar{\c}^{*}$ are the
anti-superfields corresponding to $V,~\bb,~\bar{\bb},~\rb,~\bar{\rb},~\c,~\bar{\c}$ respectively.
The ghosts
$\rb,~\bar{\rb},~\c,~\bar{\c}$ are $\s$-closed, thus their conjugate anti-superfields drop out of the BV
superspace action. In order to maintain the $\s$ BRST symmetry of the BV action  we assign to all the
anti-superfields trivial $\s$ transformations ($\s$-closed):
\begin{equation}
   \s\bm{V}^{*}=\s\bb^{*}=\s\bar{\bb}^{*}=\s\rb^{*}=\s\bar{\rb}^{*}=\s\c^{*}=\s\bar{\c}^{*}=0~.
\end{equation}
Now we can define a full superspace antibracket $\bm{(}~.~,~.~\bm{)}$
\begin{IEEEeqnarray*}{ll}
    \bm{(}\bm{F},\bm{G}\bm{)}&\equiv\int d^4x~d^2\th~d^2\thd
    \Big\{\frac{\delta \bm{F}}{\delta\bm{\Phi}^{*}_{i}}\frac{\delta \bm{G}}{\delta\bm{\Phi}^{i}}+
    \frac{\delta \bm{G}}{\delta\bm{\Phi}^{*}_{i}}\frac{\delta \bm{F}}{\delta\bm{\Phi}^{i}}\Big\}\n\\[2mm]
                             &=\int d^4x~d^2\th~d^2\thd
\Big\{~\frac{\delta \bm{F}}{\delta\bm{V}^{*}}\frac{\delta \bm{G}}{\delta V}+
    \frac{\delta \bm{G}}{\delta\bm{V}^{*}}\frac{\delta \bm{F}}{\delta V}\\[1mm]
&\hspace{3.1cm}+\frac{\delta \bm{F}}{\delta\bm{\b}^{*}}\frac{\delta \bm{G}}{\delta\bm{\b}}+
    \frac{\delta \bm{G}}{\delta\bm{\b}^{*}}\frac{\delta \bm{F}}{\delta\bm{\b}}
+\frac{\delta \bm{F}}{\delta\bar{\bb}^{*}}\frac{\delta \bm{G}}{\delta\bar{\bb}}+
    \frac{\delta \bm{G}}{\delta\bar{\bb}^{*}}\frac{\delta \bm{F}}{\delta\bar{\bb}}\\[1mm]
&\hspace{3.1cm}+\frac{\delta \bm{F}}{\delta\bm{\r}^{*}}\frac{\delta \bm{G}}{\delta\bm{\r}}+
    \frac{\delta \bm{G}}{\delta\bm{\r}^{*}}\frac{\delta \bm{F}}{\delta\bm{\r}}
+\frac{\delta \bm{F}}{\delta\bar{\rb}^{*}}\frac{\delta \bm{G}}{\delta\bar{\rb}}+
    \frac{\delta \bm{G}}{\delta\bar{\rb}^{*}}\frac{\delta \bm{F}}{\delta\bar{\rb}}\\[1mm]
&\hspace{3.1cm}+\frac{\delta \bm{F}}{\delta\bm{\c}^{*}}\frac{\delta \bm{G}}{\delta\bm{\c}}+
    \frac{\delta \bm{G}}{\delta\bm{\c}^{*}}\frac{\delta \bm{F}}{\delta\bm{\c}}
+\frac{\delta \bm{F}}{\delta\bar{\c}^{*}}\frac{\delta \bm{G}}{\delta\bar{\c}}+
    \frac{\delta \bm{G}}{\delta\bar{\c}^{*}}\frac{\delta \bm{F}}{\delta\bar{\c}}~\Big\}
\end{IEEEeqnarray*}
Using this bracket, the $\s$ invariance of the BV superspace action takes the form of a superspace classical
master equation
\begin{equation}
    \bm{(}S,S\bm{)}=2\s S=0~.
\end{equation}
As discussed previously, by populating the first slot of the antibracket with the superspace BV action $S$,
we define a nilpotent superspace BRST differential operator $\S$ in the space of superfields and anti-superfields
\begin{equation}
    \S\bm{F}\equiv\bm{(}S,\bm{F}\bm{)}~.
\end{equation}
The action of $\S$ on the superfields and anti-superfields of the vector supermultiplet is:
\begin{IEEEeqnarray*}{ll}\n\label{ss}
    \S V=\Dd^2\c+\D^2\bar{\c}~,~~&\S\bm{V}^{*}=\D^\gamma\Dd^2\D_{\gamma}V+\D^2\rb+\Dd^2\bar{\rb}~,\sn\\
    \S \rb=0~,~~&\S\rb^{*}=\bb^{*}+\D^2V+\xi\bar{\rb}~,\sn\\
    \S\bar{\rb}=0~,~~&\S\bar{\rb}^{*}=\bar{\bb}^{*}+\Dd^2V+\xi\rb~,\sn\\
    \S \bb=\rb~,~~&\S\bb^{*}=-\D^2\Dd^2\c~,\sn\\
    \S\bar{\bb}=\bar{\rb}~,~~&\S\bar{\bb}^{*}=-\Dd^2\D^2\bar{\c}~,\sn\\
    \S \c=0~,~~&\S\c^{*}=-\Dd^2\bm{V}^{*}+\Dd^2\D^2\bb~,\sn\\
    \S\bar{\c}=0~,~~&\S\bar{\c}^{*}=-\D^2\bm{V}^{*}+\D^2\Dd^2\bar{\bb}~.\sn
    \end{IEEEeqnarray*}
As expected the action of $\S$ on superfields is identical to their $\s$ transformations, while $\S$ on
anti-superfields gives the equations of motion for the corresponding superfields. By splitting these equations
of motion into two pieces, the piece coming from the anti-superfield terms of the action and the rest we can
decompose $\S$ into two nilpotent and anticommuting operators $\g$ and $\d$: $\S=\g+\d$. The $\g$ part is the
superspace BRST symmetry operator and $\d$ is the superspace Koszul-Tate resolution differential:
\begin{IEEEeqnarray*}{llll}\n
    \g V=\Dd^2\c+\D^2\bar{\c}~,~~&\g\bm{V}^{*}=0~,~&
    \d V=0~,~~&\d\bm{V}^{*}=\D^\gamma\Dd^2\D_{\gamma}V+\D^2\rb+\Dd^2\bar{\rb}~,\sn\\
    \g \rb=0~,~~&\g\rb^{*}=\bb^{*}~,~&
    \d \rb=0~,~~&\d\rb^{*}=\D^2V+\xi\bar{\rb}~,\sn\\
    \g\bar{\rb}=0~,~~&\g\bar{\rb}^{*}=\bar{\bb}^{*}~,~&
    \d\bar{\rb}=0~,~~&\d\bar{\rb}^{*}=\Dd^2V+\xi\rb~,\sn\\
    \g \bb=\rb~,~~&\g\bb^{*}=0~,~&
    \d \bb=0~,~~&\d\bb^{*}=-\D^2\Dd^2\c~,\sn\\
    \g\bar{\bb}=\bar{\rb}~,~~&\g\bar{\bb}^{*}=0~,~&
    \d\bar{\bb}=0~,~~&\d\bar{\bb}^{*}=-\Dd^2\D^2\bar{\c}~,\sn\\
    \g \c=0~,~~&\g\c^{*}=-\Dd^2\bm{V}^{*}~,~&
    \d \c=0~,~~&\d\c^{*}=\Dd^2\D^2\bb~,\sn\\
    \g\bar{\c}=0~,~~&\g\bar{\c}^{*}=-\D^2\bm{V}^{*}~,~&
    \d\bar{\c}=0~,~~&\d\bar{\c}^{*}=\D^2\Dd^2\bar{\bb}~.\sn
\end{IEEEeqnarray*}
Using the above it is straightforward to verify that $\g^2=0,~\d^2=0,~\{\g,\d\}=0$.

The last step, is to use the $\S$ transformations in order to define a nilpotent BRST charge operator in a
Hilbert space of states. In the previous section, we demonstrated explicitly how this can be done for the
Maxwell theory. The result of that procedure can be summarized in four steps. First, we take as an input the
list of operators that appear in the nilpotent differential operator defined by the antibracket.
Second, if some of these operators carry free indices
then we introduce pairs of creation and annihilation oscillators with appropriate statistics to
dress them and absorb these free indices. Third, we calculate the algebra of all these dressed operators and
apply the Fradkin and Fradkina algorithm \cite{Fradkin:1977xi} for the construction of a corresponding
Hilbert space, nilpotent BRST charge operator.
Finally, we select an appropriate vacuum state in order to
define the cohomology of the BRST charge in this Hilbert space.

In this case, according to equations \eqref{ss} we must consider the following list of superspace operators
\begin{equation}
    \Big\{~\D^{\gamma}\Dd^2\D_{\gamma},~\D^2,~\Dd^2,~\D^2\Dd^2,~\Dd^2\D^2\Big\}
\end{equation}
None of these operators carry free indices, therefore we do not require the introduction of any additional
oscillators besides the ghost oscillators of the Fradkin-Fradkina process. The algebra of these operators is
as follows:
\begin{IEEEeqnarray*}{l}\n
    [\D^{\gamma}\Dd^2\D_{\gamma},\D^2]=0~,~[\D^{\gamma}\Dd^2\D_{\gamma},\Dd^2]=0~,~
    [\D^{\gamma}\Dd^2\D_{\gamma},\D^2\Dd^2]=0~,~[\D^{\gamma}\Dd^2\D_{\gamma},\Dd^2\D^2]=0~,\sn\\[2mm]
    [\D^2,\Dd^2]=\D^2\Dd^2-\Dd^2\D^2~,~[\D^2,\D^2\Dd^2]=-\Box~\D^2~,~
    [\D^2,\Dd^2\D^2]=\Box~\D^2~,~\sn\\[2mm]
    [\Dd^2,\D^2\Dd^2]=\Box~\Dd^2~,~[\Dd^2,\Dd^2\D^2]=-\Box~\Dd^2~,~\sn\\[2mm]
    [\D^2\Dd^2,\Dd^2\D^2]=0~\sn~.
\end{IEEEeqnarray*}
Notice that the d'Alembertian operator emerges in some of the commutators. In superspace, this is not an
independent operator because it can be expressed as a linear combination of other operators
\footnote{We follow the conventions of \emph{Superspace}}:
\begin{equation}\label{box}
    \Box=\D^2\Dd^2+\Dd^2\D^2-\D^{\gamma}\Dd^2\D_{\gamma}~.
\end{equation}
This makes the above algebra non-linear. The construction of BRST operators of
non-linear algebras has been found in\cite{Schoutens:1989tn, Buchbinder:2007au}. Using these results, we
find that the most general BRST charge we can write is:
\begin{equation}
    \Q=\eta~\D^{\gamma}\Dd^2\D_{\gamma}+\bar{\zeta}~\D^2+\zeta~\Dd^2+\xi~\D^2\Dd^2+\bar{\xi}~\Dd^2\D^2
    +\kappa~\Box-\bar{\zeta}\zeta\pi_{\xi}+\bar{\zeta}\zeta\pi_{\bar{\xi}}
\end{equation}
where $\kappa=-\xi\bar{\zeta}\pi_{\bar{\zeta}}-\bar{\zeta}\bar{\xi}\pi_{\bar{\zeta}}
-\zeta\xi\pi_{\zeta}-\bar{\xi}\zeta\pi_{\zeta}$ and $\eta,~\zeta,~\bar{\zeta},~\xi,~\bar{\xi}$ are fermionic
ghost oscillators with conjugate oscillators
$\pi_{\eta},~\pi_{\zeta},~\pi_{\bar{\zeta}},~\pi_{\xi},~\pi_{\bar{\xi}}$.
The $\D^2\Dd^2$ and $\Dd^2\D^2$ terms in the above expression can be decomposed into an anticommutator
and a commutator
\begin{equation}
    \D^2\Dd^2=\tfrac{1}{2}\{\D^2,\Dd^2\}+\tfrac{1}{2}[\D^2,\Dd^2]~,~
    \Dd^2\D^2=\tfrac{1}{2}\{\D^2,\Dd^2\}-\tfrac{1}{2}[\D^2,\Dd^2]
\end{equation}
and the BRST charge takes the form
\begin{equation}
    \Q=\eta~\D^{\gamma}\Dd^2\D_{\gamma}+\bar{\zeta}~\D^2+\zeta~\Dd^2+\r~[\D^2,\Dd^2]
    +2(\bar{\zeta}\r\pi_{\bar{\zeta}}-\zeta\r\pi_{\zeta})~\Box-\bar{\zeta}\zeta\pi_{\r}+\b~\{\D^2,\Dd^2\}
\end{equation}
where $\b=\tfrac{1}{2}(\xi+\bar{\xi}),~\r=\tfrac{1}{2}(\xi-\bar{\xi})$ and
$\pi_{\r}=\pi_{\xi}-\pi_{\bar{\xi}}$ is $\r$'s conjugate oscillator
$\{\r,\pi_{\r}\}=1$.
Because of \eqref{box}, the $\{\D^2,\Dd^2\}$ term is not an independent generator and its effect is already
captured by the $\D^{\g}\Dd^2\D_{\g}$ and $\Box$ terms. Therefore, we can ignore this term (choose $\b=0$) and focus on the cohomology of the following BRST
charge
\begin{equation}\label{sQ}
    \Q=\eta~\D^{\gamma}\Dd^2\D_{\gamma}+\bar{\zeta}~\D^2+\zeta~\Dd^2+\r~[\D^2,\Dd^2]
    +2(\bar{\zeta}\r\pi_{\bar{\zeta}}-\zeta\r\pi_{\zeta})~\Box-\bar{\zeta}\zeta\pi_{\r}
\end{equation}
acting in the reduced Hilbert space generated out of the fermionic oscillators $\eta,~\zeta,~\bar{\zeta},~\r$
and their conjugates.
In this Hilbert space, we select the vacuum
$|0\rangle$ such that it is annihilated by the oscillators $\pi_{\eta},~\bar{\zeta},~\zeta,~\r$
\begin{equation}
    \pi_{\eta}|0\rangle=\bar{\zeta}|0\rangle=\zeta|0\rangle=\r|0\rangle=0~.
\end{equation}
Thus the most general state of this Hilbert space takes the form
\begin{equation}
    |\bm{\Phi}\rangle=\sum_{a,b,c,d}\eta^{a}\pi_{\r}^{b}\pi_{\bar{\zeta}}^{c}\pi_{\zeta}^{d}|\bm{\Phi}_{a,b,c,d}\rangle
\end{equation}
where $a,b,c,d=0,1$, $|\bm{\Phi}_{a,b,c,d}\rangle=\bm{\Phi}_{a,b,c,d}|0\rangle$ and $\bm{\Phi}_{\a,b,c,d}$
is a superfield
coefficient. The Hilbert space ghost number of this state is:
\begin{equation}
    gh(|\bm{\Phi}\rangle)=a-b-c-d~.
\end{equation}
The zero ghost state (physical state) is
\begin{equation}
    |\bm{\Psi}\rangle=|\bm{V}\rangle+\eta\pi_{\r}|\bm{A}\rangle+\eta\pi_{\bar{\zeta}}|\bm{B}\rangle
    +\eta\pi_{\zeta}|\bar{\bm{B}}\rangle
\end{equation}
and we can also construct three gauge parameter states with ghost values -1 ($|\bm{\Lambda}_{-1}\rangle$),~-2
($|\bm{\Lambda}_{-2}\rangle$) and -3 ($|\bm{\Lambda}_{-3}\rangle$):
\begin{IEEEeqnarray*}{l}\n
   |\bm{\Lambda}_{-1}\rangle=\pi_{\r}|\bm{\lambda}\rangle+\pi_{\bar{\zeta}}|\bm{L}\rangle
   +\pi_{\zeta}|\bar{\bm{L}}\rangle+\eta\pi_{\r}\pi_{\bar{\zeta}}|\bm{\omega}_1\rangle
   +\eta\pi_{\r}\pi_{\zeta}|\bm{\omega}_2\rangle+\eta\pi_{\bar{\zeta}}\pi_{\zeta}|\bm{\omega}_3\rangle~,\sn\\
   |\bm{\Lambda}_{-2}\rangle=\pi_{\r}\pi_{\bar{\zeta}}|\bm{w}_1\rangle+\pi_{\r}\pi_{\zeta}|\bm{w}_2\rangle
 +\pi_{\bar{\zeta}}\pi_{\zeta}|\bm{w}_3\rangle+\eta\pi_{\r}\pi_{\bar{\zeta}}\pi_{\zeta}|\bm{w}_4\rangle~,\sn\\
 |\bm{\Lambda}_{-3}\rangle=\pi_{\r}\pi_{\bar{\zeta}}\pi_{\zeta}|\bm{z}\rangle~.\sn
\end{IEEEeqnarray*}
The transformation laws of the superfield coefficients at each ghost level are derived by the action
of the BRST charge \eqref{sQ} on the previous level state. Hence we get:
\begin{equation}
\delta_{\bm{Q}}|\bm{\Lambda}_{-2}\rangle=\Q|\bm{\Lambda}_{-3}\rangle~\Rightarrow~
    \begin{cases}
       \delta|\bm{w}_1\rangle=\Dd^2|\bm{z}\rangle~,\\
       \delta|\bm{w}_2\rangle=-\D^2|\bm{z}\rangle~,\\
       \delta|\bm{w}_3\rangle=[\D^2,\Dd^2]|\bm{z}\rangle~,\\
       \delta|\bm{w}_4\rangle=\D^{\gamma}\Dd^2\D_{\gamma} |\bm{z}\rangle~
    \end{cases}
\end{equation}
\begin{equation}
\delta_{\bm{Q}}|\bm{\Lambda}_{-1}\rangle=\Q|\bm{\Lambda}_{-2}\rangle~\Rightarrow~
    \begin{cases}
       \delta|\bm{\lambda}\rangle=-\D^2|\bm{w}_1\rangle-\Dd^2|\bm{w}_2\rangle+|\bm{w}_3\rangle~,\\
       \delta|\bm{L}\rangle=-\Dd^2|\bm{w}_3\rangle+[\D^2,\Dd^2]|\bm{w}_1\rangle+2\Box|\bm{w}_1\rangle~,\\
       \delta|\bar{\bm{L}}\rangle=\D^2|\bm{w}_3\rangle+[\D^2,\Dd^2]|\bm{w}_2\rangle-2\Box|\bm{w}_2\rangle~,\\
       \delta|\bm{\omega}_1\rangle=\D^{\gamma}\Dd^2\D_{\gamma} |\bm{w}_1\rangle-\Dd^2|\bm{w}_4\rangle~,\\
       \delta|\bm{\omega}_2\rangle=\D^{\gamma}\Dd^2\D_{\gamma} |\bm{w}_2\rangle+\D^2|\bm{w}_4\rangle~,\\
       \delta|\bm{\omega}_3\rangle=\D^{\gamma}\Dd^2\D_{\gamma} |\bm{w}_3\rangle-[\D^2,\Dd^2]|\bm{w}_4\rangle
    \end{cases}
\end{equation}
\begin{equation}
\delta_{\bm{Q}}|\bm{\Psi}\rangle=\Q|\bm{\Lambda}_{-1}\rangle~\Rightarrow~
    \begin{cases}
       \delta|\bm{V}\rangle=\D^2|\bm{L}\rangle+\Dd^2|\bar{\bm{L}}\rangle+[\D^2,\Dd^2]|\bm{\lambda}\rangle~,\\
       \delta|\bm{A}\rangle=\D^{\gamma}\Dd^2\D_{\gamma} |\bm{\lambda}\rangle+\D^2|\bm{\omega}_1\rangle
       +\Dd^2|\bm{\omega}_2\rangle-|\bm{\omega}_3\rangle~,\\
       \delta|\bm{B}\rangle=\D^{\gamma}\Dd^2\D_{\gamma} |\bm{L}\rangle+\Dd^2|\bm{\omega}_3\rangle
       -[\D^2,\Dd^2]|\bm{\omega}_1\rangle-2\Box|\bm{\omega}_1\rangle~,\\
       \delta|\bar{\bm{B}}\rangle=\D^{\gamma}\Dd^2\D_{\gamma} |\bar{\bm{L}}\rangle-\D^2|\bm{\omega}_3\rangle
       -[\D^2,\Dd^2]|\bm{\omega}_2\rangle+2\Box|\bm{\omega}_2\rangle~\\
    \end{cases}
\end{equation}
The equation of motion for the physical state is:
\begin{equation}
    \Q|\bm{\Psi}\rangle=0~\Rightarrow~\D^{\gamma}\Dd^2\D_{\gamma}|\bm{V}\rangle-\D^2|\bm{B}\rangle
    -\Dd^2|\bm{\Gamma}\rangle-[\D^2,\Dd^2]|\bm{A}\rangle=0~.
\end{equation}
Using the gauge freedom, the states $|\bm{A}\rangle$ and $|\bm{\lambda}\rangle$ can be eliminate and we get
\begin{equation}\label{Veq}
    \D^{\gamma}\Dd^2\D_{\gamma}|\bm{V}\rangle=|\bm{\Phi}\rangle+|\bar{\bm{\Phi}}\rangle
\end{equation}
where $|\bm{\Phi}\rangle$ is the chiral state $|\bm{\Phi}\rangle$=$\Dd^2|\bar{\bm{B}}\rangle,~
\Dd_{\ad}|\bm{\Phi}\rangle$=$0$
and $|\bar{\bm{\Phi}}\rangle$ is the antichiral state $|\bar{\bm{\Phi}}\rangle$=$\D^2|\bm{B}\rangle,~
\D_{\a}|\bar{\bm{\Phi}}\rangle$=$0$. The transformation of the states
$|\bm{V}\rangle,~|\bm{\Phi}\rangle,~|\bar{\bm{\Phi}}\rangle$ are
\begin{equation}
    \delta|\bm{V}\rangle=\D^2|\bar{\bm{L}}\rangle+\Dd^2|\bm{L}\rangle~,~\delta|\bm{\Phi}\rangle=0,~
    \delta|\bar{\bm{\Phi}}\rangle=0
\end{equation}
and as an integrability condition to \eqref{Veq}, the chiral and antichiral states satisfy their expected
equations of motion
\begin{equation}
    \D^2|\bm{\Phi}\rangle=0~,~\Dd^2|\bar{\bm{\Phi}}\rangle=0~.
\end{equation}

This includes a consistent sector of the theory where the chiral and antichiral state vanish
($|\bm{\Phi}\rangle=0=|\bar{\bm{\Phi}}\rangle$). This sector correspond to
the free, massless vector supermultiplet given by the state $|\bm{V}\rangle$ which satisfies the equation of motion
\begin{equation}
\D^{\gamma}\Dd^2\D_{\gamma}|\bm{V}\rangle=0~,~\delta|\bm{V}\rangle=\D^2|\bar{\bm{L}}\rangle+\Dd^2|\bm{L}\rangle~.
\end{equation}
as expected from the superspace action \eqref{V}.
\section{Superspace BRST description of Super Yang Mills}
The full $4D,~\N=1$ SYM theory, which is the non-abelian extension of the vector multiplet, is
known to be described by the superspace action
\begin{equation}
    S=-\tfrac{1}{4}~Tr\int~d^4x~d^4\th~\big(e^{-V}\D^{\g}~e^{V}\big)\Dd^2\big(e^{-V}\D_{\g}~e^{V}\big)~+h.c.
\end{equation}
where $V=V^{I}T_{I}$ and $T_{I}$ are the generators of an internal symmetry group.
The action is invariant under the gauge transformation
\begin{IEEEeqnarray*}{l}\n
   e^{V'}=e^{i\bar{\Lambda}}e^{V}e^{-i\Lambda}~\Rightarrow~\delta
V=-\frac{i}{2}L_{V}\big[\Lambda+\bar{\Lambda}+
coth\big(\tfrac{1}{2}L_{V}\big)\big(\Lambda-\bar{\Lambda}\big)\big]\\
\hspace{4.4cm}=\D^2\bar{L}+\Dd^2L-\tfrac{1}{2}\big[V,\D^2\bar{L}-\Dd^2L\big]+\mathcal{O}(V^2)
\end{IEEEeqnarray*}
where $i\bar{\Lambda}=\D^2\bar{L}~,~i\Lambda=-\Dd^2L$ and $L=L^{I}T_{I}$ and $\bar{L}=\bar{L}^{I}T_{I}$.

The procedure discussed in the previous section can also be applied to this theory. The superspace $\s$ BRST
transformation is obtained by fermionizing the above gauge transformation
\begin{equation}
    \s V=\D^2\bar{\bm{c}}+\Dd^2\bm{c}-\tfrac{1}{2}[V,\D^2\bar{\bm{c}}-\Dd^2\bm{c}]+...
\end{equation}
where $\bm{c}=\bm{c}^{I}T_{I}$ and $\bar{\bm{c}}=\bar{\bm{c}}^{I}T_{I}$. The nilpotence of $\s$ on $V$
fixes the action of $\s$ on the ghosts $\bm{c}$ and $\bar{\bm{c}}$:
\begin{equation}
    \s^2V=0~\Rightarrow~\s\bm{c}=-\tfrac{1}{2}\big\{\bm{c},\Dd^2\bm{c}\big\}~,~
    \s\bar{\bm{c}}=\tfrac{1}{2}\big\{\bar{\bm{c}},\D^2\bar{\bm{c}}\big\}
\end{equation}
Using these transformations, the $\s$-exact deformation of the SYM Lagrangian is
\begin{IEEEeqnarray*}{l}\label{SYMsuperaomega}\n
    \s\bm{\Omega}=Tr\Bigg\{\rb(\D^2V+\tfrac{\xi}{2}\bar{\rb})+\bar{\rb}(\Dd^2V+\tfrac{\xi}{2}\rb)\\
    \hspace{2cm}-\bb~\D^2\Big\{-\frac{i}{2}L_{V}\big[i~\Dd^2\bm{c}-i~\D^2\bar{\bm{c}}+
coth\big(\tfrac{1}{2}L_{V}\big)\big(i~\Dd^2\bm{c}+i~\D^2\bar{\bm{c}}\big)\big]\Big\}\\
\hspace{2cm}-\bar{\bb}~\Dd^2\Big\{-\frac{i}{2}L_{V}\big[i~\Dd^2\bm{c}-i~\D^2\bar{\bm{c}}+
coth\big(\tfrac{1}{2}L_{V}\big)\big(i~\Dd^2\bm{c}+i~\D^2\bar{\bm{c}}\big)\big]\Big\}\Bigg\}
\end{IEEEeqnarray*}
where $(\rb,~\bb)$ and $(\bar{\rb},~\bar{\bb})$ are the two algebra valued BRST doublets:
$\s\bb=\rb~,~\s\rb=0~,~\\ \s\bar{\bb}=\bar{\rb}~,~\s\bar{\rb}=0$.

By adding the corresponding anti-superfields, the superspace BV Lagrangian for SYM takes the form
\begin{IEEEeqnarray*}{ll}\n
    \mathcal{L}_{BV}=Tr\Bigg\{-\tfrac{1}{4}~\big(e^{-V}\D^{\g}~e^{V}\big)\Dd^2\big(e^{-V}\D_{\g}~e^{V}\big)~
    -\tfrac{1}{4}~\big(e^{-V}\Dd^{\gd}~e^{V}\big)\D^2\big(e^{-V}\Dd_{\gd}~e^{V}\big)\\
\hspace{2.3cm}+\rb~(\D^2V+\tfrac{\xi}{2}\bar{\rb})+\bar{\rb}~(\Dd^2V+\tfrac{\xi}{2}\rb)\\
    \hspace{2.3cm}-\bb~\D^2\Big\{-\frac{i}{2}L_{V}\big[i~\Dd^2\bm{c}-i~\D^2\bar{\bm{c}}+
coth\big(\tfrac{1}{2}L_{V}\big)\big(i~\Dd^2\bm{c}+i~\D^2\bar{\bm{c}}\big)\big]\Big\}\\
\hspace{2.3cm}-\bar{\bb}~\Dd^2\Big\{-\frac{i}{2}L_{V}\big[i~\Dd^2\bm{c}-i~\D^2\bar{\bm{c}}+
coth\big(\tfrac{1}{2}L_{V}\big)\big(i~\Dd^2\bm{c}+i~\D^2\bar{\bm{c}}\big)\big]\Big\}\\
\hspace{2.3cm}+V^{*}~\Big\{-\frac{i}{2}L_{V}\big[i~\Dd^2\bm{c}-i~\D^2\bar{\bm{c}}+
coth\big(\tfrac{1}{2}L_{V}\big)\big(i~\Dd^2\bm{c}+i~\D^2\bar{\bm{c}}\big)\big]\Big\}\\
\hspace{2.3cm}-\tfrac{1}{2}~\bm{c}^{*}~\big\{\bm{c},\Dd^2\bm{c}\big\}
+\tfrac{1}{2}~\bar{\bm{c}}^{*}~\big\{\bar{\bm{c}},\Dd^2\bar{\bm{c}}\big\}
+\bb^{*}\rb+\bar{\bb}^{*}\bar{\rb}
    \Bigg\}~.
\end{IEEEeqnarray*}
After expanding around the free theory of the previous section and taking the trace the BV Lagrangian
becomes
\begin{IEEEeqnarray*}{ll}
\mathcal{L}_{BV}=&~\tfrac{1}{2}~V^{I}\D^{\g}\Dd^2\D_{\g}V_{I}
+\tfrac{1}{4}~V^{I}\D^{\g}\Dd^2\big(\D_{\g}V^{K}~V^{\Lambda}\big)f_{K\Lambda I}
+\tfrac{1}{4}~V^{I}\Dd^{\gd}\D^2\big(\Dd_{\gd}V^{K}~V^{\Lambda}\big)f_{K\Lambda I}\\
&+\rb^{I}~(\D^2V_{I}+\tfrac{\xi}{2}\bar{\rb}_{I})+\bar{\rb}^{I}~(\Dd^2V_{I}+\tfrac{\xi}{2}\rb_{I})\\
&-\bb^{I}~\D^2\Dd^2\bm{c}_{I}+\tfrac{1}{2}~\bb^{I}~\D^2\Big(V^{K}(\D^2\bar{\c}^{\Lambda}-\Dd^2\c^{\Lambda})\Big)
f_{K\Lambda I}\\
&-\bar{\bb}^{I}~\Dd^2\D^2\bar{\bm{c}}_{I}+\tfrac{1}{2}~\bar{\bb}^{I}~\Dd^2
\Big(V^{K}(\D^2\bar{\c}^{\Lambda}-\Dd^2\c^{\Lambda})\Big)f_{K\Lambda I}\\
&+V^{*I}~\Dd^2\c_{I}+V^{*I}~\D^2\bar{\c}_{I}-\tfrac{1}{2}V^{*I}~V^{K}\big(\D^2\bar{\c}^{\Lambda}
-\Dd^2\c^{\Lambda}\big)f_{K\Lambda I}\\
&-\tfrac{1}{2}~\c^{*I}~\c^{K}~\Dd^2\c^{\Lambda}f_{K\Lambda I}
+\tfrac{1}{2}~\bar{\c}^{*I}~\bar{\c}^{K}~\Dd^2\bar{\c}^{\Lambda}f_{K\Lambda I}
+\bb^{*I}~\rb_{I}+\bar{\bb}^{*I}~\bar{\rb}_{I}\\
&+\dots
\end{IEEEeqnarray*}
where $f_{IJ}{}^{K}$ are the structure constants of the internal Lie algebra
$[T_{I},T_{J}]=f_{IJ}{}^{K}T_{K}$. The algebra indices are lowered using the Cartan-Killing
metric $g_{IJ}= Tr(T_{I}T_{J})$. The constants $f_{IJK}=f_{IJ}{}^{\Lambda}g_{\Lambda K}$ have the properties
$f_{IJK}=-f_{JIK}=-f_{IKJ}$.

Using the superspace antibracket, we find
the action of $\S$ on the superfields and anti-superfields of SYM:
\begin{IEEEeqnarray*}{l}\n\label{sSYM}
\S V^{I}=\Dd^2\c^{I}+\D^2\bar{\c}^{I}
    -\tfrac{1}{2}~V^{K}(\D^2\bar{\c}^{\Lambda}-\Dd^2\c^{\Lambda})f_{K\Lambda}{}^{I}~,~~\sn\\[2mm]
\S\bm{V}^{*}{}_{I}=\D^\gamma\Dd^2\D_{\gamma}V_{I}
+\tfrac{1}{4}~\Big[~\D^{\g}\Dd^2\big(~\D_{\g}V^{K}~V^{\Lambda}\big)
+\D^{\g}\big(~\Dd^2\D_{\g}V^{K}~V^{\Lambda}\big)
-\Dd^2\D^{\g}V^{K}~\D_{\g}V^{\Lambda}\Big]f_{K\Lambda I}~~~~~~\sn\\
\hspace{1.4cm}
+\tfrac{1}{4}~\Big[~\Dd^{\gd}\D^2\big(~\Dd_{\gd}V^{K}~V^{\Lambda}\big)
+\Dd^{\gd}\big(~\D^2\Dd_{\gd}V^{K}~V^{\Lambda}\big)
-\D^2\Dd^{\gd}V^{K}~\Dd_{\gd}V^{\Lambda}\Big]f_{K\Lambda I}
+\D^2\rb_{I}+\Dd^2\bar{\rb}_{I}\\
\hspace{1.4cm}
+\tfrac{1}{2}~\Big[~\D^2\bb^{K}+\Dd^2\bar{\bb}^{K}-\bm{V}^{*K}\Big]
(\D^2\bar{\c}^{\Lambda}-\Dd^2\c^{\Lambda})f_{I\Lambda K}~,\\[2mm]
\S \rb^{I}=0~,~~\S\rb^{*}{}_{I}=\bb^{*}{}_{I}+\D^2V_{I}+\xi\bar{\rb}_{I}~,\sn\\[2mm]
\S\bar{\rb}^{I}=0~,~~\S\bar{\rb}^{*}{}_{I}=\bar{\bb}^{*}{}_{I}+\Dd^2V_{I}+\xi\rb_{I}~,\sn\\[2mm]
\S \bb^{I}=\rb^{I}~,~~\S\bb^{*}{}_{I}=-\D^2\Dd^2\c_{I}
+\tfrac{1}{2}~\D^2\Big(~V^{K}(\D^2\bar{\c}^{\Lambda}-\Dd^2\c^{\Lambda})\Big)f_{K\Lambda I}~,\sn\\[2mm]
\S\bar{\bb}^{I}=\bar{\rb}^{I}~,~~\S\bar{\bb}^{*}{}_{I}=-\Dd^2\D^2\bar{\c}_{I}
+\tfrac{1}{2}~\Dd^2\Big(~V^{K}(\D^2\bar{\c}^{\Lambda}-\Dd^2\c^{\Lambda})\Big)f_{K\Lambda I}~,\sn\\[2mm]
\S \c^{I}=-\tfrac{1}{2}~\c^{K}~\Dd^2\c^{\Lambda}~f_{K\Lambda}{}^{I}~,~~\sn\\[2mm]
\S\c^{*}{}_{I}=\Dd^2\D^2\bb{}_{I}-\Dd^2\bm{V}^{*}{}_{I}
+\tfrac{1}{2}~\Dd^2\Big[~(\D^2\bb^{K}+\Dd^2\bar{\bb}^{K}-\bm{V}^{*K})~V^{\Lambda}\Big]f_{\Lambda IK}\sn\\
\hspace{1.4cm}
-\tfrac{1}{2}~\Big(\c^{*K}~\Dd^2\c^{\Lambda}+\Dd^2(~\c^{*K}~\c^{\Lambda})\Big)f_{I\Lambda K}~,\\[2mm]
\S\bar{\c}^{I}=\tfrac{1}{2}~\bar{\c}^{K}~\D^2\bar{\c}^{\Lambda}~f_{K\Lambda}{}^{I}~,~~\sn\\[2mm]
\S\bar{\c}^{*}{}_{I}=\D^2\Dd^2\bar{\bb}_{I}-\D^2\bm{V}^{*}_{I}
-\tfrac{1}{2}~\D^2\Big[~(\D^2\bb^{K}+\Dd^2\bar{\bb}^{K}-\bm{V}^{*K})~V^{\Lambda}\Big]f_{\Lambda IK}\sn\\
\hspace{1.4cm}
+\tfrac{1}{2}~\Big(\bar{\c}^{*K}~\Dd^2\bar{\c}^{\Lambda}+\D^2(~\bar{\c}^{*K}~\bar{\c}^{\Lambda})\Big)
f_{I\Lambda K}~.
\end{IEEEeqnarray*}
It is straight forward to decompose these transformations to their $\g$ and $\d$ components as defined in
\eqref{g} and \eqref{d}. The action of the $\g$-BRST symmetry transformation on the SYM superfields and
anti-superfields is:
\begin{IEEEeqnarray*}{l}\n\label{gSYM}
\g V^{I}=\Dd^2\c^{I}+\D^2\bar{\c}^{I}
    -\tfrac{1}{2}~V^{K}(\D^2\bar{\c}^{\Lambda}-\Dd^2\c^{\Lambda})f_{K\Lambda}{}^{I}~,~~\sn\\[2mm]
\g\bm{V}^{*}{}_{I}=
-\tfrac{1}{2}~\bm{V}^{*K} (\D^2\bar{\c}^{\Lambda}-\Dd^2\c^{\Lambda})f_{I\Lambda K}~,\sn\\[2mm]
\g \rb^{I}=0~,~~\g\rb^{*}{}_{I}=\bb^{*}{}_{I}~,\sn\\[2mm]
\g\bar{\rb}^{I}=0~,~~\g\bar{\rb}^{*}{}_{I}=\bar{\bb}^{*}{}_{I}~,\sn\\[2mm]
\g \bb^{I}=\rb^{I}~,~~\g\bb^{*}{}_{I}=0~,\sn\\[2mm]
\g\bar{\bb}^{I}=\bar{\rb}^{I}~,~~\g\bar{\bb}^{*}{}_{I}=0~,\sn\\[2mm]
\g \c^{I}=-\tfrac{1}{2}~\c^{K}~\Dd^2\c^{\Lambda}~f_{K\Lambda}{}^{I}~,~~\sn\\[2mm]
\g\c^{*}{}_{I}=-\Dd^2\bm{V}^{*}{}_{I}
-\tfrac{1}{2}~\Dd^2\Big[\bm{V}^{*K}~V^{\Lambda}\Big]f_{\Lambda IK}
-\tfrac{1}{2}~\Big(\c^{*K}~\Dd^2\c^{\Lambda}+\Dd^2(~\c^{*K}~\c^{\Lambda})\Big)f_{I\Lambda K}~,\sn\\[2mm]
\g\bar{\c}^{I}=\tfrac{1}{2}~\bar{\c}^{K}~\D^2\bar{\c}^{\Lambda}~f_{K\Lambda}{}^{I}~,~~\sn\\[2mm]
\g\bar{\c}^{*}{}_{I}=-\D^2\bm{V}^{*}_{I}
+\tfrac{1}{2}~\D^2\Big[\bm{V}^{*K}~V^{\Lambda}\Big]f_{\Lambda IK}
+\tfrac{1}{2}~\Big(\bar{\c}^{*K}~\Dd^2\bar{\c}^{\Lambda}+\D^2(~\bar{\c}^{*K}~\bar{\c}^{\Lambda})\Big)
f_{I\Lambda K}\sn
\end{IEEEeqnarray*}
and the action of the Koszul-Tate complex $\d$ on the SYM anti-superfields is:
\begin{IEEEeqnarray*}{l}\n\label{dSYM}
\d\bm{V}^{*}{}_{I}=\D^\gamma\Dd^2\D_{\gamma}V_{I}
+\tfrac{1}{4}~\Big[~\D^{\g}\Dd^2\big(~\D_{\g}V^{K}~V^{\Lambda}\big)
+\D^{\g}\big(~\Dd^2\D_{\g}V^{K}~V^{\Lambda}\big)
-\Dd^2\D^{\g}V^{K}~\D_{\g}V^{\Lambda}\Big]f_{K\Lambda I}~~~~~~\sn\\
\hspace{1.4cm}
+\tfrac{1}{4}~\Big[~\Dd^{\gd}\D^2\big(~\Dd_{\gd}V^{K}~V^{\Lambda}\big)
+\Dd^{\gd}\big(~\D^2\Dd_{\gd}V^{K}~V^{\Lambda}\big)
-\D^2\Dd^{\gd}V^{K}~\Dd_{\gd}V^{\Lambda}\Big]f_{K\Lambda I}
+\D^2\rb_{I}+\Dd^2\bar{\rb}_{I}\\
\hspace{1.4cm}
+\tfrac{1}{2}~\Big[~\D^2\bb^{K}+\Dd^2\bar{\bb}^{K}\Big]
(\D^2\bar{\c}^{\Lambda}-\Dd^2\c^{\Lambda})f_{I\Lambda K}~,\\[2mm]
\d\rb^{*}{}_{I}=\D^2V_{I}+\xi\bar{\rb}_{I}~,~~
\d\bar{\rb}^{*}{}_{I}=\Dd^2V_{I}+\xi\rb_{I}~,\sn\\[2mm]
\d\bb^{*}{}_{I}=-\D^2\Dd^2\c_{I}
+\tfrac{1}{2}~\D^2\Big(~V^{K}(\D^2\bar{\c}^{\Lambda}-\Dd^2\c^{\Lambda})\Big)f_{K\Lambda I}~,\sn\\[2mm]
\d\bar{\bb}^{*}{}_{I}=-\Dd^2\D^2\bar{\c}_{I}
+\tfrac{1}{2}~\Dd^2\Big(~V^{K}(\D^2\bar{\c}^{\Lambda}-\Dd^2\c^{\Lambda})\Big)f_{K\Lambda I}~,\sn\\[2mm]
\d\c^{*}{}_{I}=\Dd^2\D^2\bb{}_{I}
+\tfrac{1}{2}~\Dd^2\Big[~(\D^2\bb^{K}+\Dd^2\bar{\bb}^{K})~V^{\Lambda}\Big]f_{\Lambda IK}~,\sn\\
\d\bar{\c}^{*}{}_{I}=\D^2\Dd^2\bar{\bb}_{I}
-\tfrac{1}{2}~\D^2\Big[~(\D^2\bb^{K}+\Dd^2\bar{\bb}^{K})~V^{\Lambda}\Big]f_{\Lambda IK}
\end{IEEEeqnarray*}

Unlike the free theory example of section \ref{vs}, the Hilbert space description of SYM can not be captured
by a sole BRST charge $\Q$. The nonlinear terms require the introduction of a new kind of product ``$\bm{*}$''
in the superfield state space. This product will assign to a pair of superfield states ($|\bm{A}\rangle,~|\bm{B}\rangle$) a new state
$|\bm{A}\rangle\bm{*}|\bm{B}\rangle$. Hence, for interacting theories, the Lagrangian BRST symmetry
operator $\S$ corresponds to the doublet $(\Q~,~\bm{*})$ acting in an appropriately defined Hilbert space of
superfield states $\S|\bm{\Psi}\rangle=\Q|\bm{\Psi}\rangle+|\bm{\Psi}\rangle\bm{*}|\bm{\Psi}\rangle$.
This is similar to string field theory, where there is a $*$-product
and the physical spectrum of the theory is described by the equation $Q\Psi+\Psi*\Psi=0$.
\section{Summary}
It has been demonstrated repeatedly that knowing the BRST description of a gauge theory is a very powerful
tool.  In this paper we explore the BRST description of supersymmetric gauge theories while maintaining
supersymmetry manifest.
For a given superspace formulated gauge theory, we follow the usual BRST procedure and construct a nilpotent
BRST symmetry operator $(\s)$ by fermionizing the original gauge symmetry. For every gauge parameter
superfield we introduce a Fadeev-Popov ghost superfield and two additional ghost superfields which play the
role of a BRST doublet. The superspace action is deformed by an appropriate $\s$-exact term which generates
gauge fixing conditions and remove negative norm states from the physical spectrum of the theory. This
deformation is further extended by the addition of anti-superfield terms according to the BV procedure.
Using a superspace anti-bracket we define the nilpotent BRST-BV differential operator ($\S$) in superspace
and identify its decomposition to the superspace BRST symmetry transformation ($\g$) and the superspace
Koszul-Tate complex ($\d$). We apply this procedure to $4D,~\N=1$ super Maxwell theory and its non-abelian
extension super Yang-Mills. For both theories, we derive explicit expressions for all these nilpotent
operators in terms of the superspace covariant derivatives.

Moreover, for the linearized theory we explore its BRST-BFV description in terms of a nilpotent BRST charge
operator ($\Q$) acting on the Hilbert space of superfield states.  Superspace BRST charges have been
constructed previously for maximally supersymmetric theories by introducing pure spinors. The nilpotence of
these BRST charges is automatic due to the pure spinor constraint. In this paper, we follow another approach
and construct a Hilbert space BRST charge based on the algebra of the appropriate set of superspace
differential operators. Specifically we consider the set of linearly independent differential operators that
appear in $\S$. Due to the algebraic properties of the supersymmetric covariant derivative, the algebra of
these operators is nonlinear.  Nevertheless, the Fradkin-Fradkina algorithm can be appropriately
modified and applied in order to construct a manifestly supersymmetric nilpotent BRST charge without requiring
pure spinor variables. By choosing an appropriate vacuum state,
we show that its physical state cohomology generates the correct superspace equations.
For the interacting theory, the physical spectrum of the theory can not be reproduced solely in terms of
a BRST charge but additional structure must be introduced in order to capture the nonlinear terms
that correspond to the gluing of superfield states in Hilbert space.

There are several interesting future directions that we want to investigate. First of all, this
methodology must be applied to gauge theories with tensorial prepotential superfields such as
supergravity and higher spin supermultiplets in order to explore the structure of the nilpotent operators for
these theories. It is very interesting that for non-supersymmetric Maxwell theory, the physical state
cohomology \eqref{spin1}, \eqref{dspin1} is also valid for higher spin gauge theories by simply allowing the
expansion of the states in terms of bosonic oscillators that carry spacetime indices. This happens because the
gauge invariance of higher spin theories (Bianchi identities) rely on exactly the same algebra as Maxwell
theory. However, for supersymmetric higher spin theories this is not true hence, we must apply the same procedure to
theories described by higher rank gauge superfields.
Secondly, we must explore the properties of the additional product rule ($\bm{*}$) structure between
superfield states required by the BRST-BFV description of interacting gauge theories.
String interactions in string field theory offers an example of such structure.

\section*{Acknowledgments}
The work of I.~L.~B was partially supported by Ministry of Education of Russian Federation, project
FEWF-2020-0003. The work of K.~K. and S.~J.~G. is supported in part by the endowment of the Ford Foundation
Professorship of Physics at Brown University. Also K.~K. gratefully acknowledges the support of the Brown
Theoretical Physics Center.
\begin{multicols}{2}
{\small
\bibliographystyle{hephys}
\bibliography{references}
}
\end{multicols}
\end{document}